\documentclass[10pt, conference, compsocconf]{IEEEtran}
\ifCLASSINFOpdf
\else
\fi

\usepackage{epsfig}
\usepackage{amsthm}
\usepackage{amsmath,amsmath,amssymb,amsfonts}

\begin{document}
%
% paper title
% can use linebreaks \\ within to get better formatting as desired
\title{Performance Acceleration of Kernel Polynomial Method\\ Applying
Graphics Processing Units}

% author names and affiliations
% use a multiple column layout for up to two different
% affiliations

\author{\IEEEauthorblockN{Shixun Zhang and Shinichi Yamagiwa}
\IEEEauthorblockA{School of Information\\
Kochi University of Technology/\\
JST PRESTO\\
Kami, Kochi 782-8502 Japan}
\and
\IEEEauthorblockN{Masahiko Okumura and Seiji Yunoki}
\IEEEauthorblockA{Computational Condensed Matter Physics Laboratory\\
RIKEN ASI\\
Wako, Saitama, 351-0198 Japan,\\
JST CREST\\
Kawaguchi, Saitama 332--0012, Japan, and \\
Computational Materials Science Research Team\\
RIKEN AICS\\
Kobe, Hyogo, 650-0047 Japan}
}

% conference papers do not typically use \thanks and this command
% is locked out in conference mode. If really needed, such as for
% the acknowledgment of grants, issue a \IEEEoverridecommandlockouts
% after \documentclass

% for over three affiliations, or if they all won't fit within the width
% of the page, use this alternative format:
% 
%\author{\IEEEauthorblockN{Michael Shell\IEEEauthorrefmark{1},
%Homer Simpson\IEEEauthorrefmark{2},
%James Kirk\IEEEauthorrefmark{3}, 
%Montgomery Scott\IEEEauthorrefmark{3} and
%Eldon Tyrell\IEEEauthorrefmark{4}}
%\IEEEauthorblockA{\IEEEauthorrefmark{1}School of Electrical and Computer Engineering\\
%Georgia Institute of Technology,
%Atlanta, Georgia 30332--0250\\ Email: see http://www.michaelshell.org/contact.html}
%\IEEEauthorblockA{\IEEEauthorrefmark{2}Twentieth Century Fox, Springfield, USA\\
%Email: homer@thesimpsons.com}
%\IEEEauthorblockA{\IEEEauthorrefmark{3}Starfleet Academy, San Francisco, California 96678-2391\\
%Telephone: (800) 555--1212, Fax: (888) 555--1212}
%\IEEEauthorblockA{\IEEEauthorrefmark{4}Tyrell Inc., 123 Replicant Street, Los Angeles, California 90210--4321}}

% use for special paper notices
%\IEEEspecialpapernotice{(Invited Paper)}

% make the title area
\maketitle

\begin{abstract}
The Kernel Polynomial Method (KPM) is one of the fast diagonalization methods used for simulations of quantum systems in research fields 
of condensed matter physics and chemistry. The algorithm has a difficulty to be parallelized on a cluster computer or a supercomputer due to the fine-gain recursive calculations. This paper proposes an implementation of the KPM on the recent graphics processing units (GPU) where the recursive calculations are able to be parallelized in the massively parallel environment. This paper also illustrates performance evaluations regarding the cases when the actual simulation parameters are applied, the one for increased intensive calculations and the one for increased amount of memory usage. Finally, it concludes that the performance on GPU promises very high performance compared to the one on CPU and reduces the overall simulation time.

\end{abstract}

\begin{IEEEkeywords}
GPGPU, Kernel Polynomial Method, Condensed Matter Physics, CUDA
\end{IEEEkeywords}

% For peer review papers, you can put extra information on the cover
% page as needed:
% \ifCLASSOPTIONpeerreview
% \begin{center} \bfseries EDICS Category: 3-BBND \end{center}
% \fi
%
% For peerreview papers, this IEEEtran command inserts a page break and
% creates the second title. It will be ignored for other modes.
\IEEEpeerreviewmaketitle

\section{Introduction}

Today's technological achievement in our everyday life is based on
years of fundamental research for a wide variety of materials with
fascinating functionalities such as semiconductors, magnets, and
superconductors.  Researchers in condensed matter physics revealed long
ago that those different properties of materials result from different
behaviors of electrons, which are described by quantum mechanical
equation of motion.  Although it has been more than 80 years since
quantum mechanics was established, there are still many properties of
matters whose origins are yet to be understood. Such examples include
copper based high temperature superconductors~\cite{Muller} and some
of magnetic insulators of organic compounds~\cite{Kato}.  The common
feature of these systems is a strong quantum correlation between
electrons, which is turned out to be crucial for determining their
properties. It is precisely this strong correlation that makes it
difficult to treat these systems analytically without introducing any
bias in theory.

The best way to treat the strong quantum correlations is to solve
quantum mechanical equation of motion numerically exactly.  Because of
exponential increase of degrees of freedom with the number of
electrons $\sim O(10^{23})$,
%applying theories in the quantum physics to the models used in the
%simulation, it is not easy to perform a perfect simulation due to an
%explosion of problem size where the number of degrees of freedom
%exponentially increases typically following $O(10^{23})$, with
%respect to the number of electrons. Therefore, it is common to
%perform the simulation using approximate numerical methods that
%evaluate physical quantities.
we must still resort some sort of approximations. However, unlike
analytical treatments, numerical simulations can handle the strong
correlation effects with controllable approximations.  Among many,
well established numerical methods thus far are exact diagonalization
method~\cite{Dagotto,Yamada}, quantum Monte Carlo method~\cite{QMC},
density-matrix renormalization group
method~\cite{WhitePRL,WhitePRB,DMRG,dexDMRG}, and kernel polynomial
method (KPM)~\cite{KPM}. Each method is suited to particular sets of
problems and at the same time each has severe limitations.
%The simulation performs
%selectively a method considered its merits and demerits by the
%researcher depending on what he/she wants to know precisely. 
For instance, the exact diagonalization method is able to evaluate the
ground state (and low energy excited states) in high accuracy, but it
is limited to a small size of systems.

The simulation evaluates various physical quantities such as density
of states (DoS) and Green's functions for electrons, which are
necessary to study electronic structures. In particular, a
straightforward method to calculate the DoS by diagonalizing a
Hamiltonian matrix requires computational complexity $O(D^3)$, where
$D$ is the system size.  This complexity is a performance bottleneck
to evaluate higher energy excited states. In this respect, the KPM has
an exceptional advantage because the KPM reduces the complexity of
diagonalization to $O(D)$ at most by truncating polynomial expansions,
which in turn controls the accuracy of the approximation.  Thus, this
paper focuses on the KPM which appropriately evaluates the DoS and
Green's function including higher energy excited states~\cite{KPM}.

The KPM is an approximation method based on polynomial expansions from
which physical quantities are evaluated.  In particular, the Chebyshev
expansion is the most common and useful polynomial to be applied.
%to the polynomial expansion method and it is widely used for the
%analysis of the strong quantum correlation of electrons.  However, it
%is likely to observe the Gibbs phenomenon in the method.
To avoid the Gibbs phenomenon due to truncated polynomial expansions
with a finite order, modified kernel polynomials are preferably used.
%the KPM improves this drawback applying suitable kernels. 
For example, the Dirac's delta function is well approximated by
truncating Chebyshev expansion with the Jackson kernel~\cite{KPM}.
Moreover, in quantum statistical mechanics, it is required to evaluate
the trace of large-dimensional Hamiltonian matrices.  This trace is
efficiently approximated by using random vectors~\cite{KPM} (we call
it ``stochastic trace method'' in this paper).  Therefore, combining
these two methods, truncated polynomial expansions and random vector
bases, allows us to evaluate the DoS and other physical quantities
with significantly reduced complexity.
%due to the truncation and the randomness, which is defined by the
%trace of Dirac's delta function with a Hamiltonian matrix argument.

%Even if the truncation in the KPM reduces the complexity, 
The computational cost inevitably increases with system sizes
considered, and with the number of polynomials kept and random vectors
generated to meet the desired accuracy.
%of the target system to be simulated explodes because the number of
%the electrons is linearly increased following the system
%size. Moreover, the computational cost is increased if we need the
%high accuracy approximation increasing the numbers of the polynomials
%and the random vectors in the KPM.
It is therefore expected to reduce the simulation latency drastically
by implementing the KPM in parallel platform.

Regarding computer hardware, the graphics processing units (GPU) have
become available to be used for acceleration platform as a substitute
of CPU. This is due to the recent drastic performance growth of
GPU. The recent GPU has already achieved the performance up to TFLOPS
order. Therefore, it is applied to various scientific fields to solve
the grand challenge applications under a personal computing
environment~\cite{GPGPU_survey}.

The program on GPU is called stream-based program which processes each
data unit contained in input data streams, and generates the
corresponding data unit forming output data streams. This computing
style has benefits of 1) eliminating memory access bottleneck, which
is seen in the von Neumann style architecture, and 2) data parallelism
because each data unit does not have any dependency in the data
streams. The recent challenges to speedup intensive computations 
enforce algorithms to be redesigned to fit to GPU and to receive the 
benefit of especially the data parallelism characteristics assigning small 
operations to the enormous number of the stream processors. 
This computing style would become a typical computing style in the next
supercomputing generation. 

This paper focuses on a GPU-based implementation of the KPM applying
the stream-based computing style. We propose an effective
implementation of the KPM on GPU to accelerate its performance faster
than the recent CPU. As seen in the next section, vectors (higher
order polynomials) are generated recursively. This characteristic is
suffered to parallelize the KPM effectively in a CPU-based large
system. Applying GPU resources and a stream-based programming style,
this paper will challenge to overcome the performance limitation
caused by the recursive operation.

This paper is organized as follows. Section~\ref{sec_background} 
describes the detailed explanation of KPM and the overview of the 
general purpose computing on GPU. Section~\ref{sec_KPM} proposes the 
design and implementation of KPM on GPU. Section~\ref{sec_evaluation} 
analyzes the performances of typical sets of input parameters 
used in condensed matter physics and discusses the program behaviors 
when the parameters change to increase resource usage regarding
processor and memory. Finally, section~\ref{sec_conclusions} 
concludes this paper.

\section{Background and definitions}
\label{sec_background}

\subsection{Kernel polynomial method}
\label{sec_KPM_intro}

\subsubsection{Definition}

The basis of KPM is the following (Chebyshev) polynomial expansion of
a function $f(x)$ defined in $[-1,1]$, 
\begin{equation}
f(x) = \frac{1}{\pi \sqrt{1-x^2}} \left[ \mu_0 + 2 \sum_{n=1}^{\infty}
\mu_n T_n (x) \right] \, ,  
\end{equation}
where 
\begin{equation}
\mu_n = \int_{-1}^1 \! dx \, f(x) T_n (x) \, , \label{mu}
\end{equation}
and $T_n (x)$ is the Chebyshev polynomial defined as
\begin{equation}
T_n (x) = \cos\left[ n \arccos (x) \right] \, .
\end{equation}
It should be mentioned that the Chebyshev polynomials satisfies the
following recursion relations,
\begin{align}
T_0 (x) & = 1 \, , \quad T_1 (x) = x \, , \label{T01} \\
T_{n+2} (x) & = 2 x T_{n+1} (x) - T_n (x) \, . \label{Tn}
\end{align}
KPM is defined as
\begin{equation}
f_{\rm KPM} (x) = \frac{1}{\pi \sqrt{1-x^2}} \left[ g_0 \mu_0 + 2
 \sum_{n=1}^{N-1} g_n \mu_n T_n (x) \right] \, , \label{fKPM}
\end{equation}
where the additional coefficients $g_n$ given by a kernel which
satisfies the limit
\begin{equation}
| \! | f - f_{\rm KPM} | \! | \xrightarrow{N \rightarrow \infty}{} 0
 \, , 
\end{equation}
where $| \! | \cdot | \! |$ is suitable well-defined norm.

\subsubsection{Application to quantum systems}

In quantum physics, we need to expand functions of the Hamiltonian
matrix. In this paper, we focus on the density of state (DoS). Then,
we show an example of application of KPM for calculation of DoS.

We consider the system described by the Hamiltonian matrix $H$. First,
we apply the following linear transformation in order to fit the
spectrum of $H$ to $[-1,1]$, 
\begin{equation}
\tilde{H} = (H - \alpha_+) / \alpha_- \, , \label{tilde}
\end{equation}
where 
\begin{equation}
\alpha_{\pm} = (E_{\rm upper} \pm E_{\rm lower}) / 2 \, , 
\end{equation}
The parameters $E_{\rm upper}$ and $E_{\rm lower}$ are the upper and
lower limits of the eigenvalues of $H$ obtained by the Gerschgorin
theorem. 

The density of state (DoS) $\rho (\omega)$ of the $D$-dimensional
Hamiltonian matrix $H$ is defined by
\begin{equation}
\rho (\omega) = \frac{1}{D} \sum_{k=0}^{D-1} \delta (\omega - E_k) \,
 , 
\end{equation}
where $E_k$ is the $k$-th eigenvalue and $\delta (x)$ is the delta
function. We apply the linear transformation (\ref{tilde}) and obtain
the equation
\begin{equation}
\rho (\tilde{\omega}) = \frac{1}{D} \sum_{k=0}^{D-1} \delta
 (\tilde{\omega} - \tilde{E}_k ) \, , 
\end{equation}
where
\begin{equation}
\tilde{\omega} = (\omega - \alpha_+) / \alpha_- \, . 
\end{equation}
In order to obtain the approximated DoS using KPM, the coefficients
$\mu_n$ (\ref{mu}) in this case is obtained as
\begin{align}
\mu_n & = \int_{-1}^1 \! d \tilde{\omega} \, \rho (\tilde{\omega}) T_n
 (\tilde{\omega}) \nonumber \\ 
 & = \frac{1}{D} \sum_{k=0}^{D-1} T_n (\tilde{E}_k) \nonumber \\
 & = \frac{1}{D} \sum_{k=0}^{D-1} \langle k | T_n (\tilde{H}) | k
 \rangle 
 = \frac{1}{D} {\rm Tr} [T_n (\tilde{H})] \, , \label{DoStr}
\end{align}
where $| k \rangle$ is the $k$-th eigenvector and $\langle k | = | k
\rangle^\dag$. 

\subsubsection{Stochastic evaluation of traces}

In order to evaluate the trace in Eq.(\ref{DoStr}), we introduce the
stochastic evaluation method of traces, which estimates $\mu_n$ by
average over only a small number $R \ll D$ of randomly chosen vector.

First, we introduce an arbitrary basis $\{ | i \rangle \}$ a set of
independent identically distributed random variables $\{ \xi_{r,i} |
\xi_{r,i} \in \mathbb{R} \}$ which in terms of the statistical average
$\langle \! \langle \cdot \rangle \! \rangle$ fulfill
\begin{equation}
\langle \! \langle \xi_{r,i} \rangle \! \rangle = 0 \, , \quad 
\langle \! \langle \xi_{r,i} \xi_{r',i'} \rangle \! \rangle =
 \delta_{rr'} \delta_{ii'} \, , 
\end{equation}
a random vector is defined through
\begin{equation}
| r \rangle = \sum_{i=0}^{D-1} \xi_{r,i} | i \rangle \, . 
\end{equation}
Using them, we can approximately evaluate the trace as follows,
\begin{align}
\mu_n & = \frac{1}{D} {\rm Tr} \left[ T_n (\tilde{H}) \right]
 \nonumber \\
& = \frac{1}{D} \sum_{i=0}^{D-1} \left[ T_n (\tilde{H}) \right]_{ii}
 \nonumber \\ 
& \simeq \frac{1}{D} \frac{1}{R} \sum_{i,j=0}^{D-1} \sum_{r=0}^{R-1}
 \langle \! \langle \xi_{r,i} \xi_{r,j} \rangle \! \rangle \left[ T_n
 (\tilde{H}) \right]_{ij} \nonumber \\  
& = \biggl\langle \! \! \! \biggl\langle \frac{1}{D} \frac{1}{R}
 \sum_{r=0}^{R-1} \langle r | T_n (\tilde{H}) | r \rangle
 \biggr\rangle \! \! \! \biggr\rangle \, .  
\end{align}

In order to make $\langle r | T_n (\tilde{H}) | r \rangle$, we use the
following recursive relations for the vectors $|r_n \rangle := T_n
(\tilde{H}) | r \rangle$ derived from the relations (\ref{T01}) and
(\ref{Tn}), 
\begin{align}
| r_0 \rangle & = | r \rangle \, , \quad | r_1 \rangle = \tilde{H} |
 r_0 \rangle \, , \\
| r_{n+2} \rangle & = 2 \tilde{H} | r_{n+1} \rangle - | r_n \rangle \,
 . \label{rn}
\end{align}
Then $\mu_n$ is expressed by this expression as
\begin{equation}
\mu_n \simeq \biggl\langle \! \! \! \biggl\langle \frac{1}{D}
\frac{1}{R} \sum_{r=0}^{R-1} \langle r_0 | r_n \rangle
\biggr\rangle \! \! \! \biggr\rangle \, . \label{munr}
\end{equation}

\subsubsection{Numerical complexity}

The numerical complexity of the KPM is $O(SRND)$ if the $\tilde{H}$ is
sparse matrix, where $S$ is the number of the realization of the set
of random variables $\{ \xi_{r,i} \}$. The process costing $O(D)$ is
the making part of $| r_n \rangle$ shown in Eq.~(\ref{rn}), which is
the heaviest part in KPM. When the $\tilde{H}$ is considered as a
dense matrix, the complexity of the part becomes $O(D^2)$.The $O(SR)$
comes from the average and summation in Eq.~({\ref{munr}}) and $O(N)$
from the summation in Eq.~(\ref{fKPM}). This numerical cost $O(SRND)$
is very effective against the full diagonalization which costs
$O(D^3)$ if $S,R,N \ll D^2$, and the $\tilde{H}$ is a sparse
matrix. However, it is a dense matrix, the numerical cost becomes
$O(SRND^2)$ due to all multiplications for all elements in the
$\tilde{H}$ and the $| r_n \rangle$ must be performed straightly
without considering the CRS (Compressed Row Storage) format for a
sparse matrix. This paper considers the simple case when the CRS
format is not applied to the memory maintenance for the
$\tilde{H}$. Therefore, all the elements in the $\tilde{H}$ matrix are
applied to all the calculations in the KPM.

\subsection{General purpose computing on GPUs}

\subsubsection{GPU architecture}

A video adapter that includes a GPU and a Video RAM (VRAM) is
connected to a CPU's peripheral bus such as PCI Express. The video
adapter works as a peripheral device of the CPU, and its GPU is
controlled by the CPU to help a part of visualization tasks in the
system. To utilize the GPU as a computing resource for GPGPU
applications, the CPU downloads application program to the GPU's
instruction memory and also prepares input data for the program. The
program fetches the data and generates the result to the memory
areas. The GPU reads/writes the VRAM directly to execute the
calculation for the program. In this case, the original data is
prepared in the main memory. The CPU copies the data to the
VRAM. During the execution of the program, the GPU generates the
results to the VRAM. The CPU copies the results from the VRAM to the
main memory.

The recent GPUs have only a kind of processor called the {\it stream
  processor}. The processor works for general purpose processes in any
kind of calculation. However, the computing style must be followed in
the stream-based one distributing elements included in streams into
multiple stream processors. GPU uses two types of memory called {\it
  global} and {\it shared} memories. The global memory is provided by
the memory placed outside of GPU such as DDR3 VRAM. The shared memory
is placed besides of the stream processor that works as if a cache.

\begin{figure}[t]
\centering
\epsfig{file=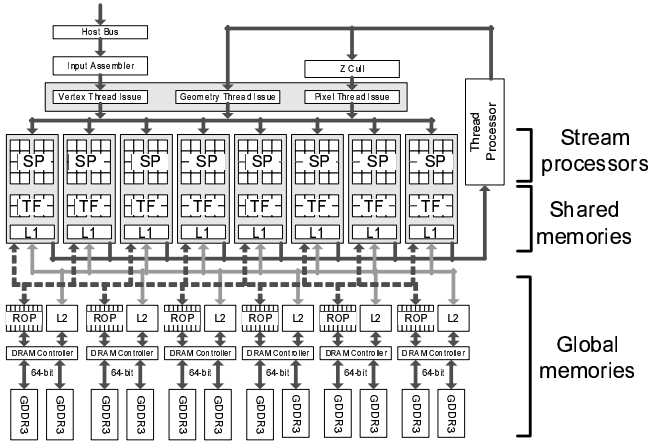, width=\linewidth}
\caption{A GPU architecture.}
\label{fig_gpu_arch}
\end{figure}

\subsubsection{CUDA}

\begin{figure}[t]
\centering
\epsfig{file=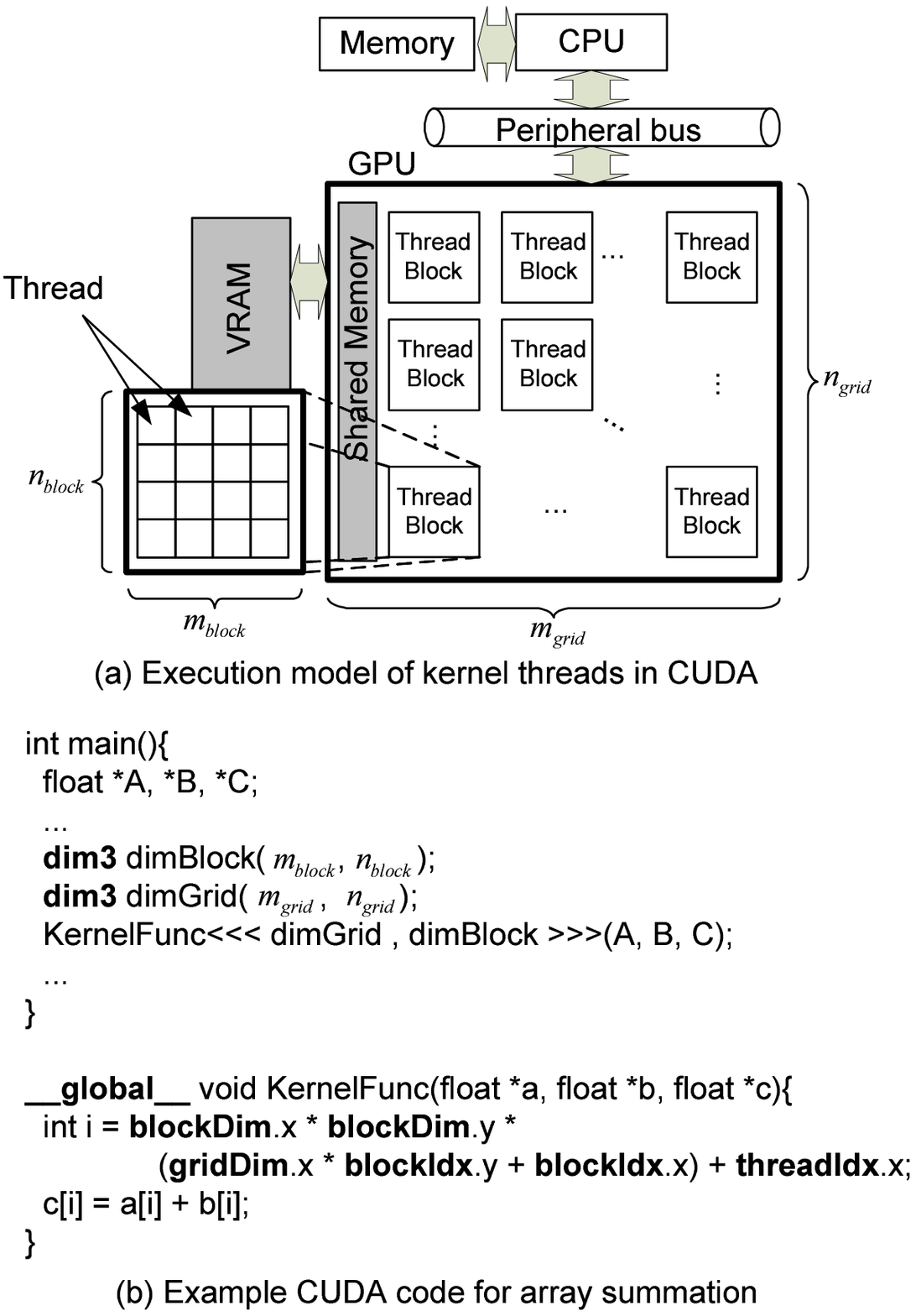, width=\linewidth}
\caption{CUDA programming environment.}
\label{fig_cuda}
\end{figure}

The Compute Unified Device Architecture (CUDA) has been proposed by
NVIDIA corporation~\cite{CUDA}. The tools and APIs for programming on
CUDA environment is now provided by the company's website.

The CUDA assumes an architecture model as illustrated in
Figure~\ref{fig_cuda} (a). The model defines a GPU which is connected
to a CPU's peripheral bus. A VRAM (the global memory) that maintains
data used for calculation on the GPU is connected to the GPU. The data
is copied from the host memory before the CPU commands to execute a
program on the GPU. The program is executed as a thread in a thread
block. The thread blocks are tiled in a matrix of from one to three
dimensions. In the figure, thread blocks are tiled in two dimensions
which size is $n_{grid}\times m_{grid}$. Each thread block has
multiple threads in a matrix which size is varied from one to three
dimensions. The figure also shows a thread block that includes
$n_{block}\times m_{block}$ threads. Each thread block has individual
shared memory space where shared valuables accessed among threads in
the block are stored temporally. Thus, the program targeted to GPU in
the CUDA environment is invoked as threads. The threads are grouped by
the unit of the thread block. Therefore, obtaining a large
parallelism, a large number of threads are invoked concurrently.

In the program on the CUDA environment, the threads are described as a
stream-based function written in C called a {\it kernel function} as
shown in Figure~\ref{fig_cuda} (b). The program has two parts of the
codes targeted to CPU and GPU, which is initially invoked by the CPU;
a main program for CPU and a kernel function called as the thread on
GPU. The kernel function is defined with the {\tt \_\_global\_\_}
directive so that it is executed on GPU. In the function, the global
variables named {\tt gridDim}, {\tt blockDim}, {\tt blockIdx}, {\tt
  threadIdx}, implicitly declared by the CUDA runtime, are available
to be used to specify the size of the grid and the thread block, the
indices of the thread block and of the thread respectively. For
example, using these global valuables, Figure~\ref{fig_cuda} (b)
performs a summation of arrays A and B assigning each summation of the
elements in those arrays to a thread and returns the result to the
array C. The function is called by the main program specifying the
sizes of the grid and the thread block with {\tt <<< >>>}. Finally,
reading data from the VRAM transferred by the main program, the kernel
function is assigned to GPU, and runs as multiple threads. Thus,
because programmer can just simply consider the stream-based kernel
function and the calling code for the function in the main program,
using the conventional C language manner, the CUDA provides an easy
and transparent interface for GPGPU.

According to the backgrounds we have mentioned above, it is important
for the simulation in the quantum physics to apply a fast
diagonalization method to reach the goal of the simulation
quickly. However, the KPM has difficulty of fine-grain parallelization
in large scale computers such as cluster computers or supercomputers
due to the recursive calculation performed in the
Eq.~(\ref{rn}). Therefore, it is worth for us to implement the KPM on
a GPU where the massively parallel environment is equipped with a
large number of stream processors. Thus, this paper focuses on design
and implementation of the KPM on GPU that challenges to achieve the
advanced performance with applying the stream-based computing style on
the massively parallel environment.

\section{Kernel polynomial method applying GPUs}
\label{sec_KPM}

\subsection{Design for massively parallel platform}

\begin{figure}[t]
\centering
\epsfig{file=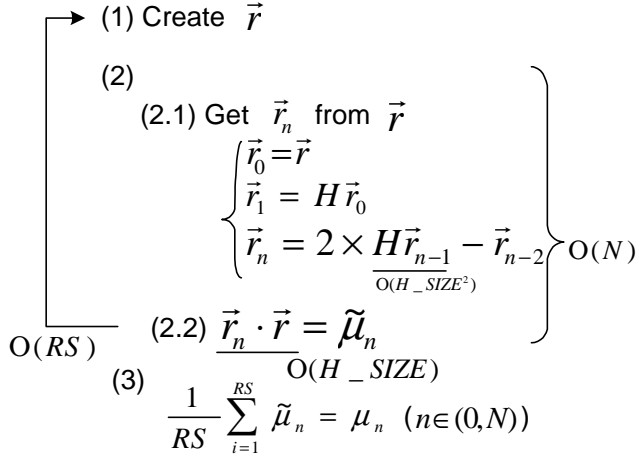, width=\linewidth}
\caption{Design and implementaion of KPM on GPUs.}
\label{fig_algorithm}
\end{figure}

Figure~\ref{fig_algorithm} summarizes the KPM algorithm. The step (1)
generates randomly a vector $\overrightarrow{r}$ that the number of
elements is $H\_SIZE$ (this equals to the $D$ in
section~\ref{sec_KPM_intro}). The step (2) gets $\overrightarrow{r}_n$
from $\overrightarrow{r}_{n-1} $ and $\overrightarrow{r}_{n-2} $
recursively calculating a matrix multiply of $H$ and
$\overrightarrow{r}_{n-1} $ in the step (2.1). This multiplication is
very hard to parallelize using MPI or OpenMP because of the
dependencies due to the recursive iteration although the part needs
the most intensive calculation. Then a dot product is calculated using
$\overrightarrow{r}_n $ again with $\overrightarrow{r} $ at the step
(2.2) and generates $\tilde{\mu}_n$. Then the generation of the
$\tilde{\mu}_n$ is iterated for $RS$ times. This means each generation
of $\tilde{\mu}_n$ can be massively parallelized on GPUs. Finally, the
average of all the $\tilde{\mu}_n$s is generated at the step (3). N
$\mu_n$s are finally generated from the RS-time iterations of the step
(1) and (2). This generation of the moments achieves the objective of
the KPM. This summation to generate $\tilde{\mu}_n$ can be
parallelized on GPU. Therefore, implemented on GPUs, two parallel
processing parts are entirely performed during the evaluation of the
moments using KPM: a) generation of $\tilde{\mu}_n$ and b) generation
of $ \mu_n$. The maximum number of parallelism at the both a) and b)
parts becomes the $SR$ because the total number of threads executed in
the stream processors is $SR$. Here, GPU has an architectural
restriction to the number of threads in a thread block referred as
BLOCK\_SIZE in this paper. Therefore, the number of thread blocks
becomes $RS/BLOCK\_SIZE$. Considering the parallelization techniques
above, let us explain the implementation of a kernel program on CUDA
that invokes both the a) and b) parts.

\subsection{Implementation}

\begin{figure}[t]
\centering
\epsfig{file=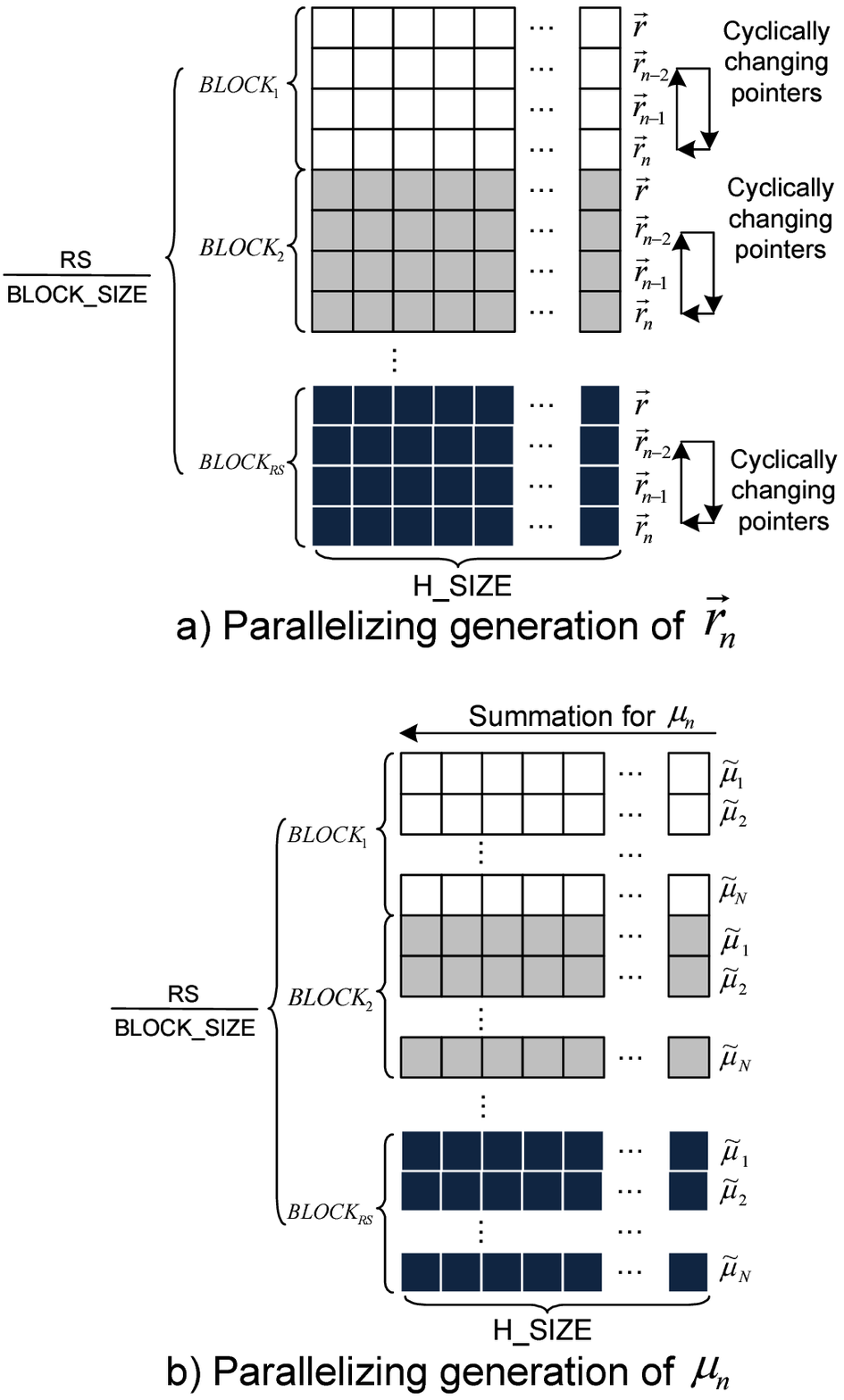, width=7cm}
\caption{Parallelization of KPM.}
\label{fig_parallelize}
\end{figure}

We have implemented a kernel program for GPU using CUDA. The kernel
receives the $H\_SIZE$, $N$ that is the number of moments and $RS$ as
the arguments. All calculations are performed based on double
precision. The kernel includes two important concepts. One is how to
keep high parallelism. Another is an effective memory management for
the parallelism.

\subsubsection{Parallelization of calculations}

As we discussed in the last section, two heavy calculation parts (the
a) and the b)) should be parallelized and it would give the largest
impact for the speedup.

Figure~\ref{fig_parallelize} a) shows the generation part for the
$\overrightarrow{r}_n$.  $\overrightarrow{r}_n$ needs
$\overrightarrow{r}_{n-1}$, $\overrightarrow{r}_{n-2}$ and
$\overrightarrow{r}$ that is randomly generated. These four vectors
are obtained in the global memory and each block will write those
vectors swapping the pointers. Here the number of blocks is
$RS/BLOCK\_SIZE$. In each block, $BLOCK\_SIZE$ stream processors are
concurrently working to generate a part of those vectors including the
random number generation for $\overrightarrow{r}$. Therefore, this
part will be fully parallelized into the total number of stream
processors equipped on GPUs. This part will generate $\tilde{\mu}_1$,
$\tilde{\mu}_2$ ... $\tilde{\mu}_N$ using $\overrightarrow{r}$ and
$\overrightarrow{r}_n$ for $N$ time iteration.

Figure~\ref{fig_parallelize} b) depicts the parallelization of
generation for $\mu_n$. It performs just parallel summations for
generating a scalar $\mu_n$ where all blocks works in parallel.

\subsubsection{Memory consumption}

Let us consider the required memory amount for the operations in
Figure~\ref{fig_parallelize} in the case of double precision.  For the
operation a), because four $\overrightarrow{r}$ vectors are stored in
the global memory for a block. Each $\overrightarrow{r}$ vector has
$H\_SIZE$ elements. Therefore, this part consumes $Number~of~Blocks
\times 4 \times H\_SIZE \times 8$ bytes. The operation b) is
parallelized into the number of blocks. Each block performs a part of
summation using N $\tilde{\mu}$s. The length of $\tilde{\mu}$s is
$H\_SIZE$. Therefore, it needs totally $Number~of~Blocks \times N
\times H\_SIZE \times 8$ bytes.

The operation a) writes $\tilde{\mu}_n$ into the global memory. This
needs to be kept with $\overrightarrow{r}$ vectors
simultaneously. Therefore, the total number of memory is
$Number~of~Blocks \times H\_SIZE \times (8\times N + 32)$.

Due to the recursive relationships among $\overrightarrow{r}_n$,
$\overrightarrow{r}_{n-1}$ and $\overrightarrow{r}_{n-2}$, the KPM is
treated generally as one of very hard parallelized
algorithms. However, as we can see in this section, on the GPU, a
massively parallel environment, the KPM is fully parallelized due to
the stream-based computing concept. Thus, we can expect an effective
speedup that will be proportional to the number of stream-processors.

\section{Experimental performance analysis}
\label{sec_evaluation}

This section shows performance evaluations of the KPM implemented on
GPU. The performance based on GPU is compared with the one based on
CPU. The experimental environment is a PC that consists of an Intel's
Core i7 930 processor at 2.80GHz with 12GB DDR3 memory, and the NVIDIA
Tesla C2050 with 3GB Memory is connected to the PCI Express bus. The
configuration of the cache in the GPU is set to 16KB and the shared
memory size is 48KB. The OS of the PC is the Cent OS of the Linux
Kernel 2.6.18. The driver version of the GPU is 3.0. All KPM
calculations are performed with double precision floating point. The
CPU version is compiled with GCC 4.4.1 with O3 option.

We perform three kinds of performance analysis: (1) evaluation using
actual sets of parameters, (2) the one with increasing calculation
size and (3) the one with increasing memory usage. The first
evaluation hires sets of parameters used in actual simulations of the
meaningful model applied to the condensed matter physics field. The
second evaluation analyses the behavior of the performances when the
parameter $N$ is increased. This means that more intensive calculation
is loaded to the CPU and the GPU following the increase of $N$. The
last evaluation shows the performance impacts when the $H\_SIZE$ is
increased. This case needs the square sized memory to store the $H$
matrix that is increased by the impact of $H\_SIZE^2$.

\subsection{Performance analysis using actual simulation parameters}

\begin{figure}[t]
\centering
\epsfig{file=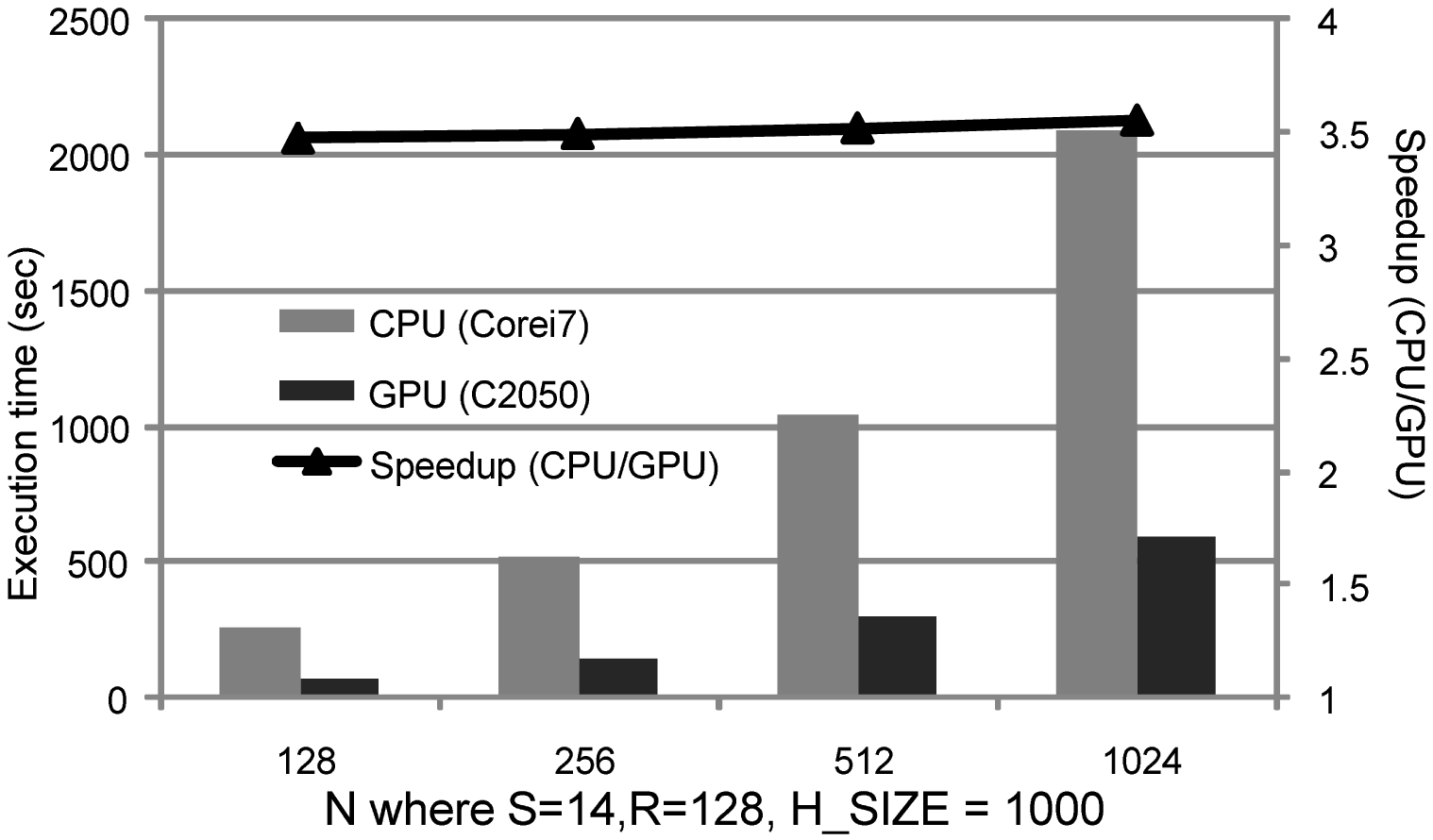, width=\linewidth}
\caption{Performances applying the lattice made of cubes placed in 10x10x10.}
\label{fig_realistic}
\end{figure}

\begin{figure}[t]
\centering
\epsfig{file=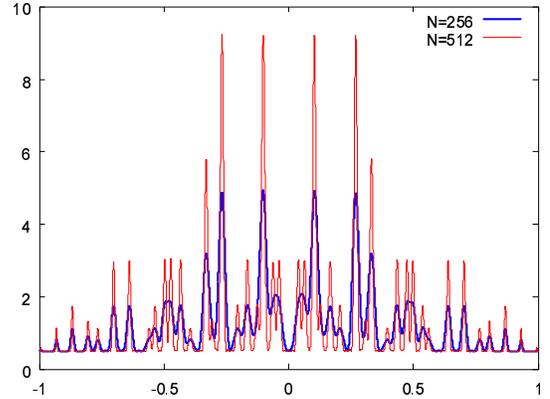, width=\linewidth}
\caption{The DoS comparison with trancations between N=256 and N=512 when the lattice is made of cubes placed in 10x10x10, R=14 and S=128.}
\label{fig_moment}
\end{figure}

In the field of the computational condensed matter physics, the KPM is
applied to a simulation to evaluate the DoS in a three dimensional
lattice model. Let us consider a lattice model made of cubes in $10
\times 10 \times 10$ where an electron is placed in each corner. This
model needs a Hamiltonian matrix sized in $1000 \times 1000$ due to
the presentations of correlations among the electrons at each
corner. The significant characteristics of the matrix include that 1)
it is sparse and symmetric and 2) any row contains seven non-zero
elements with the condition where all diagonal ones are zeros and the
other non-zero ones are $-1$s.

We evaluate the DoS in the case of the lattice that we assumed above
using the fixed parameters of the KPM with $S=14$ and $R=128$. Varying
$N$ from 128 to 1024 in the steps of $2^n$, Figure~\ref{fig_realistic}
shows the execution times and the speedup comparing the performances
on the CPU with the ones on the GPU. The speedup keeps 3.5 times for
all the cases. This means that the simulation can be accelerated by
the GPU and the execution time becomes almost 40\% faster than the one
on CPU at most.

We shall pickup two DoS data combinations from the parameter sets of
Figure~\ref{fig_realistic} and plot it to a graph as depicted in
Figure~\ref{fig_moment}. The graph shows the DoS when $N=256$ and
$N=512$. When $N$ is the smaller number, the truncation reduces to the
resolution of the DoS. However, the processing time is smaller than
the case of a large $N$. Therefore, although the case of $N=512$ shows
higher resolution of the DoS, it takes longer calculation time.

\subsection{Performance analysis with increased intensive calculations}

\begin{figure}[t]
\centering
\epsfig{file=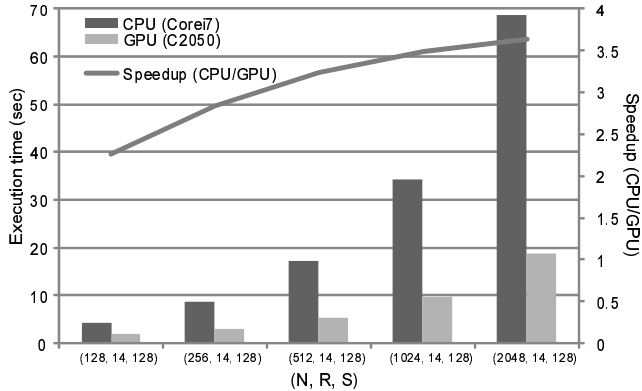, width=\linewidth}
\caption{Performance comparison increasing N.}
\label{fig_intensive}
\end{figure}

Obtaining the fixed parameters of $H\_SIZE=128$, $R=14$ and $S=128$,
we measure the performances with varying the $N$ from 128 to 2048. The
graph of the performances is illustrated in
Figure~\ref{fig_intensive}. The graph shows the execution times with
bars and the speedups (i.e. the CPU time is divided by the
corresponding GPU time) with a line. As increasing the $N$, that is,
as increasing the calculation amount, the speedup increases to almost
4 times. This means that the performance with the higher intensive
calculations affected by the larger $N$ causes higher effective data
parallelism on GPU when the calculation amount is increased without
changing the size of the memory usage. Thus, our implementation on GPU
clearly achieves higher performance than the CPU-based KPM as
increasing the calculation amount.

\subsection{Performance analysis with increased memory usage}

\begin{figure}[t]
\centering
\epsfig{file=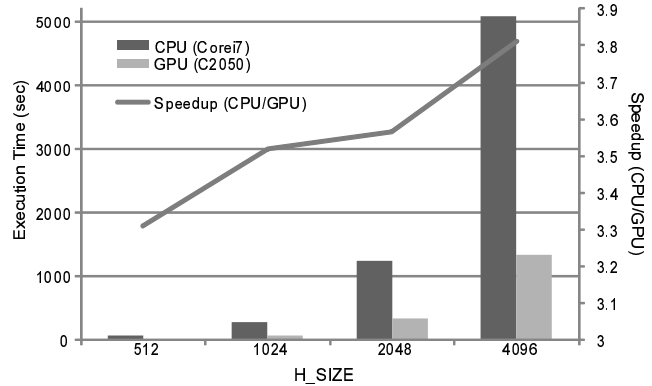, width=\linewidth}
\caption{Performance comparison increasing H\_SIZE.}
\label{fig_mem}
\end{figure}

This analysis fixes $N=128$, $R=14$ and $S=128$. We vary $H\_SIZE$
from 512 to 4096 with the step of $2^n$. The performance presents
effects caused by increasing the memory usage. The graph of the
performance is depicted in Figure~\ref{fig_mem}. When the amount of
memory usage increases, the number of memory accesses
increases. Therefore, the CPU version needs to read/write the memory
as increased the size of $\tilde{H}$ matrix. On the other hand,
because the GPU can cache a part of the matrix into very fast shared
memory and accesses the memory in the stream-based manner. Thus, the
execution time of the GPU version does not increase more than the
complexity ($O(H\_SIZE^2)$). This causes almost four times faster
performance than the CPU version.

As we discussed in three kinds of evaluations above, the performances
on GPU achieve better performances than the ones on CPU due to the
highly parallelism caused by the GPU-based implementation explained in
this paper. The implementation achieves the advanced performance even
if it is applied to the actual examples from the condensed matter
physics or the cases with hard conditions virtually when the amounts
of the computation and the memory usage are increased. Thus, we have
confirmed that the KPM is a suitable algorithm that fits well to the
GPU environment and the performance acceleration accomplishes
amazingly the high performance.

\section{Conclusions}
\label{sec_conclusions}

This paper has proposed an implementation of the KPM widely used in
the physics and the chemistry field to simulate various quantum
states. Our GPU version shows about 4 times faster than the CPU
one. Therefore, using a GPU, productivity of the moments for a quantum
state is accelerated to four times. Therefore, the GPU version is
expected to be used for various grand challenge simulations to find a
new quantum state that resolves unknown physical theories in the
natural phenomenon.

For the future plans, we are considering to quest a method to find the
best block size used in the GPU that defines the size of the stream
processors' block. Moreover, the parallelization of the KPM on a
message passing and a shared memory paradigm is also challenging
because the recursive reference to get $\overrightarrow{r}_n$ becomes
a bottleneck to be parallelized in fine-grain. Moreover, we are also
planning to extend the GPU-based implementation to a GPU cluster for
its parallelization.

% use section* for acknowledgement
\section*{Acknowledgment}

This work is partially supported by the Japan Science Technology
Agency (JST) PRESTO program.

% trigger a \newpage just before the given reference
% number - used to balance the columns on the last page
% adjust value as needed - may need to be readjusted if
% the document is modified later
%\IEEEtriggeratref{8}
% The "triggered" command can be changed if desired:
%\IEEEtriggercmd{\enlargethispage{-5in}}

% references section

% can use a bibliography generated by BibTeX as a .bbl file
% BibTeX documentation can be easily obtained at:
% http://www.ctan.org/tex-archive/biblio/bibtex/contrib/doc/
% The IEEEtran BibTeX style support page is at:
% http://www.michaelshell.org/tex/ieeetran/bibtex/
\bibliographystyle{IEEEtran}
% argument is your BibTeX string definitions and bibliography database(s)
\bibliography{APDCM11_Phys}
%
% <OR> manually copy in the resultant .bbl file
% set second argument of \begin to the number of references
% (used to reserve space for the reference number labels box)
%\begin{thebibliography}{1}
%
%\bibitem{IEEEhowto:kopka}
%H.~Kopka and P.~W. Daly, \emph{A Guide to \LaTeX}, 3rd~ed.\hskip 1em plus
%  0.5em minus 0.4em\relax Harlow, England: Addison-Wesley, 1999.
%
%\end{thebibliography}

% that's all folks
\end{document}